\documentclass[aps,prl,twocolumn,superscriptaddress,groupedaddress]{revtex4}
\usepackage{graphicx}
\usepackage{dcolumn}  
\usepackage{bm}       
\usepackage{amssymb}

\begin{document}

\title{Einstein-Podolsky-Rosen paradox in single pairs of images}

\author{Eric Lantz}
\email[Corresponding author: ]{eric.lantz@univ-fcomte.fr}
\affiliation{D\'epartement d'Optique, Institut FEMTO-ST, Universit\'e Bourgogne Franche-Comt\'e, CNRS, Besan\c{c}on, France}
\author{S\'everine Denis}
\affiliation{D\'epartement d'Optique, Institut FEMTO-ST, Universit\'e Bourgogne Franche-Comt\'e, CNRS, Besan\c{c}on, France}
\author{Paul-Antoine Moreau}
\affiliation{D\'epartement d'Optique, Institut FEMTO-ST, Universit\'e Bourgogne Franche-Comt\'e, CNRS, Besan\c{c}on, France}
\author{Fabrice Devaux}
\affiliation{D\'epartement d'Optique, Institut FEMTO-ST, Universit\'e Bourgogne Franche-Comt\'e, CNRS, Besan\c{c}on, France}

\date{\today}

\begin{abstract}
Spatially entangled twin photons provide a test of the Einstein-Podolsky-Rosen (EPR) paradox in its original form of position (image plane) versus impulsion (Fourier plane). We show that recording a single pair of images in each plane is sufficient to safely demonstrate an EPR paradox. On each pair of images, we have retrieved the fluctuations by subtracting the fitted deterministic intensity shape and then have obtained an intercorrelation peak with a sufficient signal to noise ratio to safely distinguish this peak from random fluctuations.  A $95\%$ confidence interval has been determined, confirming a high degree of paradox whatever the considered single pairs. Last, we have verified that the value of the variance of the difference between twin images is always below the quantum (poissonian) limit, in order to ensure the particle character of the demonstration. Our demonstration shows that a single image pattern can reveal the quantum and non-local behavior of light, without any need of averaging after repeating the experiment.
\end{abstract}

\maketitle

Statistical properties of fluctuations in quantum mechanics are described by ensemble averages, which are often estimated by time averages if the signal is stationary in time, but which can also be estimated by spatial averages if the signal is stationary in space on a sufficiently large area. Most of the experiments in quantum imaging record averages of temporal coincidences, i.e. characterize the spatial repartition of temporal averages, rather than spatial averages, with accent on the high dimensionality of the underlying entanglement, in order to demonstrate that an image conveys a great number of spatially correlated quantum (temporal) channels in parallel \cite{Dixon2012,Howell2004,Krenn2014}. However, patterns in an image are pure spatial information, without any time aspect, that is ultimately degraded by spatial fluctuations of quantum origin in very weak images. Studying these fluctuations involve the use of cameras and needed so far the use of several images to exhibit quantum features. The sub shot-noise nature of the correlation between twin images issued from spontaneous down-conversion has been demonstrated either for a mean of several photons per pixel with low noise charge coupled devices \cite{Brida2009,Ottavia04} or in the photon-counting regime with electron-multiplying charge coupled devices (EMCCD) \cite{Blanchet2008}. These correlations were subsequently used to improve imaging\cite{Brida2010}. In that work, the fluctuations of the idler image were subtracted from the signal image, resulting in an improvement of the image only for a degree of correlation higher than 0.5., attained for slightly more than half of the images. Hence, though the improvement was purely spatial, the use of a set of numerous images was necessary in this experiment, leaving open the question of demonstrating pure spatial quantum effects in single images, without the need of either temporal averages or averages of set of images.

The same remark can be applied to our recent demonstration of spatial Einstein-Podolsky-Rosen (EPR) paradox \cite{Moreau2014}. EPR showed \cite{Einstein1935} that quantum mechanics predicts that entangled particles could have both perfectly correlated positions and momenta, in contradiction with the so-called \textit{local realism} where two distant particles should be treated as two different systems. Though the original intention of EPR was to show that quantum mechanics is not complete, the standard present view is that entangled particles do experience nonlocal correlations \cite{Bell1964,Aspect1981,Banaszek1998}. Spontaneous down conversion (SPDC) provides independent pairs of entangled photons that makes the system very close of that considered in the original EPR paper: the positions of photons are detected in the near-field and their momenta correspond to the far-field. Howell et al \cite{Howell2004} have measured in both planes the probability distribution of the position of the second photon, conditioned by the detection of the first photon behind a slit. This type of detection selects a priori photons experiencing temporal coincidences. On the other hand, all photons were recorded with an equal chance in our experiments with one\cite{Moreau2012} or two\cite{Moreau2014} EMCCD cameras,  showing a high degree of paradox by using spatial correlation of the images. However, it was necessary in these works to record a set of couples of images, at least 20 couples in the near field, in order to, first, retrieve the fluctuations by subtracting an average image and, second, compute the intercorrelation coefficient by adding the results of each couple of images. In a related work \cite{Edgar2012}, EPR-type correlations were obtained by averaging on a set of $10^5$ images, with however an intercorrelation peak that emerges above noise for at least hundred images. Note that, because of type 1 phase matching, results were obtained in this latter work for one dimension with no spatial separation between the twin photons.

\begin{figure}[htbp]
\centering
\includegraphics[width=\linewidth]{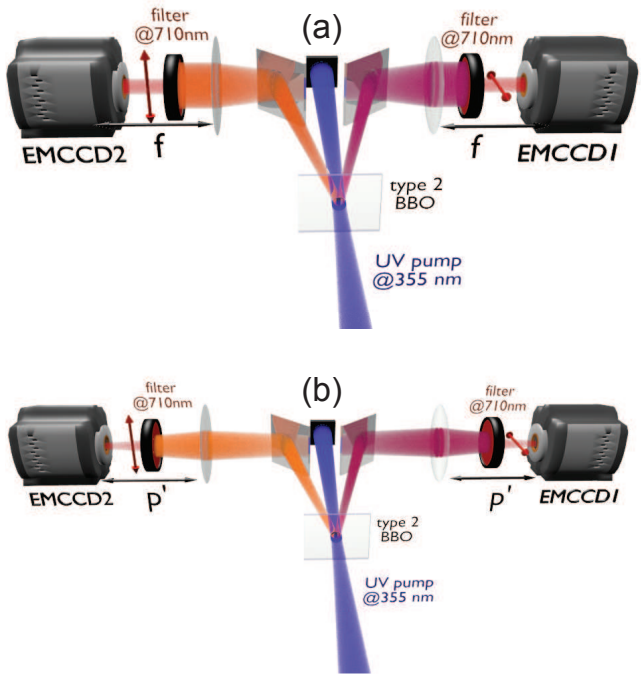}
\caption{Experimental setups used to imaging correlations. (a), measurement of momentum correlations with the cameras in the focal plane. (b), cameras in the crystal image plane.}
\label{fig1}
\end{figure}

In the present work we show that a single couple of images in each plane (near and far field) is sufficient to safely demonstrate an EPR paradox. This improvement of our previous results has been made possible by suppressing the fluorescence parasitic light coming from the elements of the experimental set-up, allowing an increase of the integral of the correlation coefficient from 0.1 to above 0.2. This integral, i.e. the degree of correlation, represents the ratio between the number of photons detected in pairs and the total number of photons, with a value of one in an ideal experiment where all photons are recorded with their twin. With this improved value, the number of coherence cells in a single image is sufficient to ensure a signal-to-noise ratio (SNR) allowing a safe distinction between the true correlation peak and random fluctuations. Indeed, we have shown in \cite{Lantz2014} that this SNR can be expressed as:
 \begin{equation}\label{SNRbin}
 \mathit{SNR}\simeq \sqrt{C K}\frac{\eta\  m}{m+p_{n}}.
 \end{equation}
 where $C$ is the number of independent coherence cells in one image, $K$ the number of recorded couples of images, $\eta$ the effective quantum efficiency of the whole system, $m$ the mean number of photons per pixel and $p_{n}$ the noise level per pixel, due either to electronics or parasitic light. Eq. \ref{SNRbin} is valid if pixels are binned in order to form "superpixels" with a size corresponding to a coherence cell. The width of such a coherence cell, or spatial mode, is proportional in the near field to the inverse of the phase-matching range in the far-field, and in the far-field to the inverse of the the pump size in the near-field~\cite{Saleh2000, Devaux2012}. Note that the number of coherence cells corresponds to the two-photon Schmidt number\cite{Exter2006}. It is possible to determine the minimum total number of coherence cells ensuring a determination without ambiguities of the quantum correlation peak. If we assume that the fluctuations of the accidental coincidences are Gaussian, they never exceed five standard deviations. For $m\gg p_{n}$, it means that no ambiguity is possible if $\mathit{SNR}>5$, or $C K>\left(\frac{5}{\eta}\right)^2$. For a single couple of images, $K=1$, and an overall quantum efficiency of 0.2, it corresponds to 25 coherence cells in each transverse direction.
 
 \begin{figure}[ht]
 	\centering
 	\includegraphics[width=6.5cm]{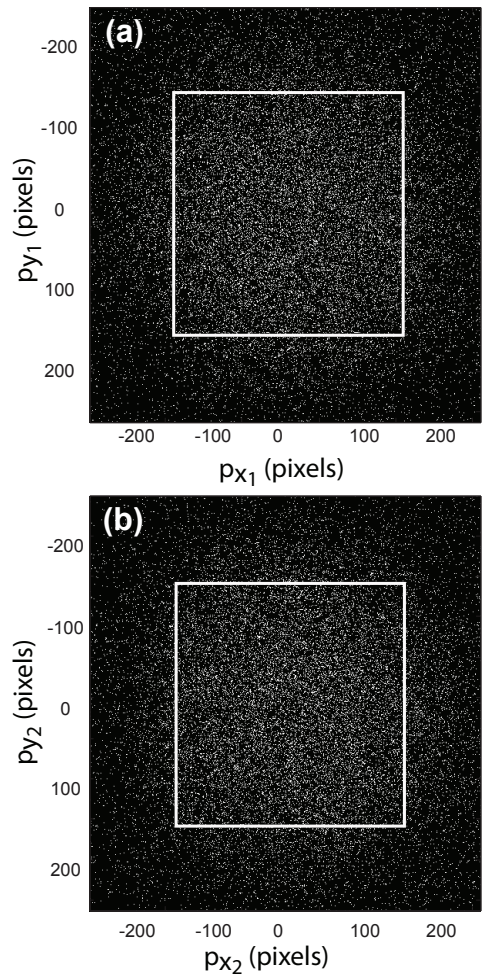}
 	\caption{Example of a pair of (a) a signal image and (b) an idler image in the far-field (focal) plane. $p_{x1}, p_{x2}, p_{y1}, p_{y2}$ are the coordinates of pixels versus the image center. Images have been thresholded and the white square lines encompass the pixels that are used for the correlation process.}
 	\label{images}
 \end{figure}

The experimental setup, shown in Fig.\ref{fig1}, is similar to that used in\cite{Moreau2014}, except that the mount of optical components no more includes polymer, that emitted parasitic fluorescent light. Pump pulses at 355 nm provided by a 27 mW laser illuminate a $0.8$-mm long $\beta$ barium borate (BBO) nonlinear crystal cut for type-II phase matching. The signal and idler photons are separated by means of two mirrors and sent to two independent imaging systems. The far-field image of the SPDC is formed on the EMCCDs placed in the focal plane of two 120-mm lenses (Fig.\ref{fig1}a). In the near-field configuration, Fig.\ref{fig1}b, the plane of the BBO crystal is imaged on the EMCCDs with a transversal magnification $M=2.44\pm0.02$. Note that only the positions of the lenses and cameras are different in the two configurations. The back-illuminated EMCCD cameras (Andor iXon3) have a quantum efficiency greater than $90\%$ in the visible range. The detector areas are formed by $512\times512$ pixels, with a pixel size of $s_{pix} = 16\times16\mu m^2$. The used images correspond to the $300\times300$ central  pixels in both the near and far-fields, including most of the SPDC photons. An image corresponds to the summation of 30 laser shots, i.e an exposure time of 0.03s.  Photon pairs emitted around the degeneracy are selected by means of narrow-band interference filters centered at 710 nm ($\Delta\lambda = 4 nm$). The photon-counting regime is ensured by adjusting the exposure time in such a way that the mean fluency of SPDC is between 0.1 and 0.2 photon per pixel in order to minimize the whole number of false detections\cite{Lantz2008}. The mean number of photons per spatiotemporal mode is less than $10^{-3}$, in good agreement with the hypothesis of pure spontaneous parametric down-conversion, without any stimulated amplification. A thresholding procedure is applied\cite{Lantz2008} to convert the gray scales into binary values that correspond to 0 or 1 photon. An example of the obtained images is given in Fig.\ref{images}.

\begin{figure}[htbp]
	\centering
	\includegraphics[width=\linewidth]{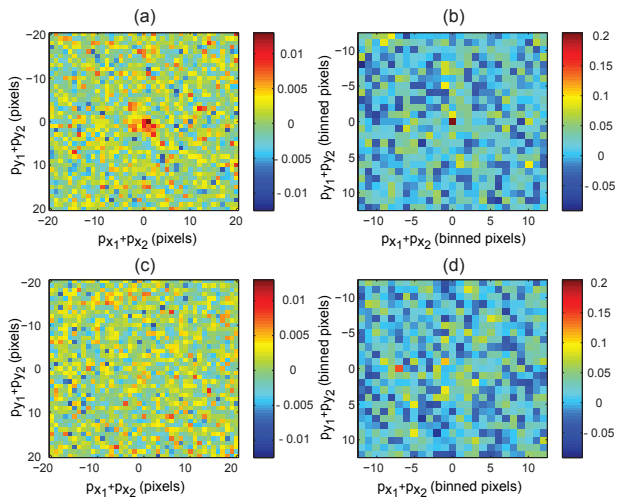}
	\caption{Cross-correlation function of the couple of twin images of Fig.\ref{images} (far-field) and comparison with decorrelated images. (a): Twin images without binning. (b): Twin images with 11$\times$11 binned pixels. (c) and (d): Decorrelated images without (c) and with (d) binning.}
	\label{Correl}
\end{figure}

Spatial correlations between the signal and the idler images have two origins. One is quantum: photons are emitted by pairs. A second, classical correlation arises from the non spatially stationary nature of the intensity: the mean intensity follows the gaussian profile of the pump beam in the near-feld, or the phase-matching profile in the far-field, leading to correlations because a photon is more likely recorded in the higher intensity part of each image. In our previous experiment\cite{Moreau2014}, we worked on the fluctuations with respect to this mean profile, obtained by subtracting from each image the sum of all images of the set. In the present experiment, we have to use only a couple of images in each plane. It is nevertheless possible to work on quantum fluctuations by subtracting a fitted deterministic gaussian (near-field) or sinc-like (far-field) profile from the image. Because of the great number of pixels of the image, the deterministic shape is correctly determined on a single image, alllowing the retrieval of quantum fluctuations by subtraction.

Fig. \ref{Correl}(a) shows an example of the obtained cross-correlation function in the far-field. The quantum correlation peak is clearly visible, with an extension that corresponds to the dimensions of a coherence cell. However, the maximum value of this peak has the same order of magnitude as random maxima due to random fluctuations. On the other hand, after binning the pixels of the cross-correlation image over a $11\times11$ pixels coherence cell, the quantum peak becomes much higher than any random peaks :  Fig.\ref{Correl}(b). We have verified that no correlation is visible for a couple signal-idler issued from different laser shots: Fig. \ref{Correl}(c) and (d). After binning, the quantum peak, identified by its position,  is the highest in 900 couples of images over 900 in the far-field, and 895 over 900 in the near-field. 

These results are consistent with the experimental and theoretical values of the SNR, as inferred from Eq.\ref{SNRbin}. First, the effective quantum efficiency $\eta$ can be  assessed as :  
\begin{equation}\label{eta}
\eta=\eta_{filter}\times \eta_{optics}\times \eta_{camera}=0.56\times 0.64\times 0.74=0.26
\end{equation}
where $\eta_{filter}$ takes into account the transmission of the interferential filter\cite{Devaux2012}, $\eta_{optics}$ the transmission of all other optical elements, including a dichroic mirror, and $\eta_{camera}$ the camera quantum efficiency as well as the probability of a false detection due to thresholding. Experimentally, the value of $\eta$ can be compared to the integral of the correlation peak, calculated with the 900 images, equal to 0.23 in the far-field and 0.19 in the near-field. The agreement is correct, meaning that parasitic light is weak, if not completely negligible. After binning, the SNR obtained by summing the correlations of the 900 images (after subtraction of their mean) has a value of 185 in the far-field, in full agreement with Eq.\ref{SNRbin}, with $C=27\times27$ coherence cells in the image. The same agreement is observed in the near field.

Let us return to the one image case. For K=1, we expect from Eq.\ref{SNRbin} a mean SNR equal to 6.2 in the far-field and to 5.1 in the near-field. We obtain experimentally a mean of 6.8 in the far-field with 95\%  of the values between 4.2 and 9.5. For the near-field, the mean is 5.3, with 95\% of the values between 2.6 and 7.9. We can conclude from these numbers that:
\begin{itemize}
\item the agreement between the results of Eq.\ref{SNRbin} for K=1 and the observed SNR is rather good. The slightly too high observed values originate from weak residual deterministic correlations due to a non perfect subtraction of the deterministic intensity shape, because of its estimation by fitting.
\item the SNR in the far-field is sufficient to safely evidence the quantum peak in all images. On the other hand, some low values of the SNR in the near-field lead to rare errors, when one of the 700 pixels of the correlation image has a value greater than the value of the quantum peak. Nevertheless, we can conclude, with 5 errors over 900 images, that we safely retrieve the quantum correlation in both the far and the near field.
\end{itemize}
To demonstrate an EPR paradox, we have to use one couple of images in each plane. We have first determined, using a gaussian fitting, the widths of the correlation peaks on each couple of images, without binning.  The results are reported in Table \ref{tab}, for the two orthogonal directions of the transverse plane x and y. 
\begin{table}[ht]
\caption{Inferred variances.}
\centering
\begin{tabular}{|l r|}
\hline
Variances & Measured values\\
\hline
$\Delta^2(x_1-x_2)\ \ $ & $\ \left[ 177\ \  931 \right] \mu m^2$ \\
$\Delta^2(y_1-y_2)\ \ $ & $\ \left[199\ \  1149\right] \mu m^2$ \\
$\Delta^2(p_{x1}+p_{x2})\ \ $ & $\ \left[  4.5 \ \ 12.3\right]\cdot 10^{-6}\hbar^2\ \mu m^{-2}$ \\
$\Delta^2(p_{y1}+p_{y2})\ \ $ & $\ \ \left[ 2.1 \ \ 5.8\right]\cdot 10^{-6}\hbar^2\ \mu m^{-2}$ \\
\hline
\end{tabular}
\label{tab}
\end{table}
Second, we aim to obtain a degree of violation for one couple of images in each plane. Because there is no reason to associate a particular couple to another, we have calculated the degree of violation for all the $900\times900$ possible combinations of two couples. By eliminating the $2,5 \%$ lowest and highest product values, we obtain the following $95 \%$ confidence intervals:
\begin{eqnarray}
0.25\hbar^2/(\Delta^2(x_1-x_2)\Delta^2(p_{x1}+p_{x2}))&=&\left[31 \ \  224\right] \\
0.25\hbar^2/(\Delta^2(y_1-y_2)\Delta^2(p_{y1}+p_{y2}))&=&\left[52 \ \ 403\right] 
\end{eqnarray}
These results exhibit a high degree of EPR paradox in the two transverse dimensions, whatever the considered couples of single images.  
\\Finally, we have verified that the binned images exhibit a sub-shot-noise statistics in both the near-field $r_{n}=0.89\pm 0.10$ and the far-field: $r_{f}=0.88\pm0.09$,  where $r$ is defined by :
\begin{equation}
r=\frac{\left\langle \Delta^2(N_1-N_2)\right\rangle}{\left\langle N_1+N_2\right\rangle}
\end{equation}
that is, the variance of the photon number difference $N_1(\boldsymbol{\rho})-N_2(\boldsymbol{\rho})$ (and $N_1(\boldsymbol{p})-N_2(-\boldsymbol{p})$ in far field) normalized to be expressed in shot noise units. $\boldsymbol\rho$ and $\boldsymbol{p}$ represent the pixel coordinates: we have employed the same coordinates system for all images, avoiding an "optimized" shift of an image with respect to its twin, that can lead  to  artificially too low values of $r$. The given confidence intervals correspond to $95\%$ of the images. Note that $r>1$ only for 8 images over 900 in the far-field and 19 images over 900 in the near-field: with an excellent confidence, the sub-shot noise character of the correlation can be shown with  a single couple of images. We have also verified that images issued from different laser shots do not exhibit such a sub-shot noise behavior. For these images, $r=1.02\pm 0.12$ both in the near-field and the far-field. 

To conclude, we have shown that a two dimensional EPR paradox can be safely demonstrated with single images, without any aspect implying temporal averaging of a set of images. Hence, an image can exhibit quantum properties by itself: all quantum correlations can be demonstrated by "repeating the experiment" over the different resolution cells, or spatial modes, of the image, and  by using  neither detection of temporal coincidences nor repeat of the experiment on a set of images.

\section*{Funding Information}
Labex ACTION program (ANR-11-LABX-0001-01)


\end{document}